\documentstyle[preprint,psfig,aps]{revtex}
\begin{document}
\draft
\title{Optical response of small silver clusters}
\author{K. Yabana\footnote{E-mail address yabana@nt.sc.niigata-u.ac.jp}}
\address{Graduate School of Science and Technology, Niigata University\\
Niigata 950-21, Japan\\
and\\}
\author{G.F. Bertsch\footnote{E-mail bertsch@phys.washington.edu}}
\address{Physics Department and Institute for Nuclear
Theory\\
University of Washington, Seattle, WA 98195 USA
}
\date{text/silver/silver3.tex; March 25, 1999}
\maketitle
\begin{abstract}
The time-dependent local density approximation is applied
to the optical response of the silver clusters, Ag$_2$,
Ag$_3$, Ag$_8$ and Ag$_9^+$.  The calculation includes all the electrons
beyond the closed-shell Ag$^{+11}$ ionic core, thus including
for the first time explicitly the filled $d$-shell in the 
response.  The excitation energy of the strong surface
plasmon near 4 eV agrees well with experiment.
The theoretical transition 
strength is quenched by a factor of 4 with respect to the pure 
$s$-electron sum rule in Ag$_8$ due to the $d$-electrons.  A 
comparable amount of
strength lies in complex states below 6 eV excitation.  The
total below 6 eV, about 50\% of the $s$ sum rule, is consistent with
published experiments.
\end{abstract}
\def\be{\begin{equation}}
\def\ee{\end{equation}} 
\section{Introduction}
The optical response of clusters of IB elements has been an
interesting theoretical
challenge:  while their chemistry is dominated by the atom's single valence
$s$-electron, the electrical properties are strongly influenced
by the nearby filled $d$-shell.  Up to now, the $d$-electrons have
been treated only implicitly by a dielectric
approximation\cite{li93}. For
example, one of the interesting phenomena that has been attributed
to the $d$-electrons is the blue shift of the surface plasmon for
small clusters\cite{ti93,kr95,se97}.  The $d$-electrons also strongly screen 
the oscillator
strength of the valence electrons, and this raises the question of 
whether the theory is consistent with the measured oscillator
strength, which are only somewhat below the full sum for the 
$s$-electrons\cite{ha92,ti92,fe93}.
  
In this work, we calculate the optical response explicitly including
the $d$-electrons, using the time-dependent local density approximation
(TDLDA).   We begin by recalling the limiting behavior
in some simple extreme models.  The first is the free-electron model including
only the $s$-electrons, as in the jellium theory.  This produces a
collective mode with all of the oscillator strength at a frequency
related to the number density $n$ by $\omega_M=\sqrt{4\pi e^2 n/3 m}$. 
At the bulk density of silver, this gives an excitation energy of
5.2 eV. The second limiting case
is the Mie theory, treating the cluster as a classical dielectric sphere.
The Mie theory in the long wavelength limit gives the
optical absorption cross section as \cite{bo93b}
\be
\sigma = {4 \pi \omega R^3\over c} {\rm Im~}{\epsilon(\omega)-1\over
\epsilon(\omega)+2}
\ee
where $R$ is the radius of the sphere and 
$\epsilon(\omega)$ is the dielectric function.  
In Fig.~1 we show the result expressed as the cross section per atom, taking 
$\epsilon(\omega)$ from ref. \cite{crc96}.
The graph also shows the integrated oscillator strength per atom,
$f_E/N = \sum_{E_i<E} f_i/N$.  We see that there is
a sharp peak at 3.5-3.6 eV, but that the oscillator strength is only 1/6
of the sum rule for $s$-electrons.  Thus the effect of the screening is to push the
$s$-electron surface plasmon down from 5.2 to 3.5 eV, together with a strong
quenching of the oscillator strength.

\section{The TDLDA method}
The details of our implementation of the TDLDA are given in ref.
\cite{ya96,ya97}.  The calculation is performed in real time, which
has the advantage that the entire response is calculated at once, and
only a Fourier transformation is needed to extract strengths of individual
excitations.  The Hamiltonian we employ is one that has been frequently
used in static calculations.  The electron-electron interaction
is treated in the local density approximation
following the prescription of
ref.~\cite{pe81}.   The ionic potential is treated in the pseudopotential
approximation keeping only the $d$- and $s$-electrons active.  The
$l$-dependent pseudopotentials were constructed according to the method
of  Troullier and Martins \cite{tr91}.  We showed in ref.\cite{ya98}
that for the atom the resulting pseudopotential is adequate to describe the 
electromagnetic
response well into the continuum, even though the sum rules become 
ambiguous\cite{ya98}.  We make one further approximation in the 
Hamiltonian, treating the nonlocality in the pseudopotential by
the method of Kleinman and Bylander\cite{kl82}.  The approximation
takes one angular momentum state as given by the radial pseudopotential
and corrects the others by adding a separable function.  A
potential problem of this method is that there may be spurious deeply
bound states in some of the partial waves\cite{go90}.  We take the local wave to
be the $s$-wave, which avoids the difficulty.  

The critical numerical parameters in the implementation of the TDLDA
on a coordinate-space mesh is the mesh spacing $\Delta x$, the shape
and size of the volume in which the electron wave functions are
calculated, and the step size $\Delta t$ of the time integration.
We use a mesh size $\Delta x$ = 0.25~\AA, which is justified in the
next section that examines atomic properties.  For the volume geometry
we take a sphere of radius 6~\AA.  From experience with the jellium
model, the collective resonance frequency of Ag$_8$ should be accurate
to 0.1 eV with this box size, and the smaller clusters will be described
even more accurately.   The last numerical parameter $\Delta t$ must
be small compared to the inverse energy scale of the Hamiltonian,
which in turn is controlled by
$\Delta x$ in our method.  We find that the integration is stable 
and accurate taking $\Delta t = 0.001$ eV$^{-1}$.  The equations are integrated
to a total time of $T= 50$ $\hbar/eV$.  The inverse of this time corresponds
to the energy resolution of the theoretical spectrum.

\section{Atomic properties}
  Before presenting the results on silver clusters, we examine the
accuracy of our three-dimensional coordinate-space numerical method
for atomic properties.  We have
considered the TDLDA treatment of IB atoms in an earlier 
publication\cite{ya98}.
There we used a spherical basis and the emphasis was on the validity of 
the pseudopotential approximation for calculating the response and its
sum rule.  Here we use those results to test the implementation
of the Kohn-Sham equations on a three-dimensional mesh, which of
course is much more inefficient than the spherical representation
for atomic systems.  Comparison of the two methods is given in Table I.
We find, with a mesh of 0.25~\AA,
that orbital energies are reproduced to an accuracy of about 0.1 eV.
The ground state configuration of the Ag atom is $d^{10}s^1$
with Kohn-Sham orbital energies of the $d$-, $s$-, and $p$-orbitals
having values -7.8, -4.6 and -0.7  eV, respectively. 
In the 3-d mesh, the lack of spherical symmetry also splits the
$d$-orbitals by about 0.1 eV.  The intrinsic limitations of the TDLDA 
on physical quantities are certainly beyond the 0.1 eV accuracy level,
so we judged the 0.25~\AA~mesh adequate for our purposes.  We also show in
the table some physical quantities of interest:
the ionization
potential, the energy of the lowest excited state, and its oscillator
strength.    Although it is tempting 
to interpret the Kohn-Sham eigenvalues
as orbital energies, it is well known that the ionization potential
is not well reproduced by the highest electron's eigenvalue.
In our case here, the negative of the $s$-orbital energy, 4.6 eV,
is quite far from the empirical 7.5 eV ionization potential.
However, the LDA does much better when the total energies of the
Ag atom and the Ag$^+$ ion are compared.  We quote this number as
`I.P.' in the table.  The next quantity we examine is the
excitation energy of the lowest excited state.  The state has
a predominant $d^{10}p^1$ character; the difference in orbital
energies is quoted as $`e_p-e_s$' in the table.  The physical
excitation energy including interaction effects is shown on the line 
$E_{p\bar s}$.  The
theoretical values are obtained from the peak position in the
Fourier transform of the TDLDA response.
We see that three-dimensional mesh agrees to 0.1 eV with the spherical basis
calculation on these energies. However, the experimental excitation energy 
is lower than theory by about 10\%; this number sets the scale of the
intrinsic limitations of the TDLDA.  In the last line, we display the 
oscillator strength associated with the transition between the ground and
excited state.  Here there is some disagreement between the spherical
results and the three-dimensional results.  This might be due to
the different treatment of the pseudopotential in the two cases.
The three-dimensional treatment used the Kleinman-Bylander method
to treat the nonlocality of the pseudopotential, while in the
spherical basis, the $l$-dependent nonlocality is treated exactly.  In any
case, the three-dimensional result is within 10\% of the empirical
value.  We also include in the table the energies associated
with the excitation of a $d$-electron to the $p$-orbital.

\section{Silver dimer and trimer}
We next examine the Ag$_2$ dimer.  We take the nuclear separation
distance at 2.612 \AA~from the calculations of ref. \cite{bo93}.  
The response averaged over directions is shown in Fig.~2.  The
$s\rightarrow p$ transition is split into two modes, a longitudinal
mode at 3.2 eV and a transverse mode at 4.9 eV.  Experimentally,
the dimer has only been studied in matrices which are subject to
environmental shifts of the order of tenths of an electron volt.
Absorption peaks have been identified at 3.0 eV and 4.7 eV which
very likely correspond to the two modes found theoretically.  In 
emission, these states are shifted somewhat lower, to 2.8 and 4.5
eV. These numbers are probably a better measure of the free cluster
energies, judging by the behavior of silver atoms in a matrix.
The lower state is strongly coupled to vibrations in the data of 
ref. \cite{sc85}, supporting the interpretation of the mode
as a longitudinal excitation.  In summary, the TDLDA reproduces the 
splitting of the longitudinal and transverse modes quite accurately, 
but the average frequency of the mode is probably too high by the same
amount that we found for the atom.  We conclude that the
interaction physics between the two atoms is reasonably described
by the TDLDA.

The picture of two nearly independent states on the two atoms is
qualitatively valid also in considering the oscillator strengths
of the transitions.   The theoretical ratio of strengths for the
two states is very close to 2:1, which is expected for the
two transverse modes compared to the single longitudinal mode.  
However, the total strength of the sharp states, 1.05 electrons, is
only 80\% of the theoretical strength for separated atoms.  
Thus a significant
fraction of strength goes to a higher spectral region.  We shall see
that much of the shift is to the region 5 eV to 6 eV, where 
experimental data is still available.

The silver trimer is predicted to have a shape of an isosceles
triangle with nearly equal sides.  There are two nearly degenerate
geometries (corresponding to the E symmetry of the equilateral
triangle) with the $^2$B state in an obtuse triangle predicted
to be lowest in most calculations.  Our calculation uses
the obtuse geometry (geometry I) of ref. \cite{bo93}.
The absorption spectrum of Ag$_3$ is shown in Fig.~3.  We see
that the absorption in the 3-5 eV region is spread out among
several states.  The more complex spectrum may be due to the
low ionization potential of Ag$_3$.  According to the
Kohn-Sham eigenvalue, the binding of the highest occupied orbital is 
3.5 eV, permitting Rydberg states in this region. There is
a quantum chemistry calculation of the spectral properties of
Ag$_3$ excitations in the visible region of the spectrum\cite{wa87}.
This calculation predicted an integrated strength below 3.5 eV
of $f_E\approx 0.6$, neglecting the screening of the $d$-electrons.
In comparison we find for the same integration limit $f_E=0.1$,
a factor of 6 smaller.

\section{Ag$_8$ and Ag$_9^+$}

We shall now see that collective features of the response become
prominent going to 8-electron clusters.
In the alkali metals, clusters with 8 valence electrons have a 
sharp collective resonance associated with a nearly spherical
cluster shape and filled shells of the delocalized orbitals.  These
systems have been modeled with the spherical jellium approximation,
and the gross features of the collective resonance are reproduced.
The IB metals are quite different from the IA alkali metals, however,
in that the occupied $d$-orbitals are close to the Fermi surface
and strongly screen the $s$-electrons.  On the experimental 
side, the studies of Ag$_8$ \cite{fe93} and Ag$_9^+$ \cite{ti93} seem
to show that the oscillator strength of the $s$-electrons is not
seriously quenched by the $d$-polarizability.  An important motivation
of our study then is to see whether the simple arguments made for
a strong $d$-screening are in fact borne out by the theory treating
the $d$-electrons on an equal footing.  

There are two competing geometries in eight-atom clusters of $s$-electron
elements, having T$_d$ and D$_{2d}$ symmetry.  We have calculated the response
of both geometries, taking the bond lengths from ref.\cite{bo93}.  
The optical absorption strength function is shown in Fig.~4.
Also shown with arrows are the two experimental absorption peaks 
seen in ref. \cite{fe93}.  The peak locations agree very well with
the theoretical spectrum based on the T$_d$ geometry. But one should remember that
the matrix spectrum is likely to be shifted by a few tenths of an eV
with respect to the free cluster spectrum.  The experimental absorption
strength is considerably higher for the upper of the two peaks in
the 3-4 eV region, which also agrees with theory.  The D$_{2d}$ 
geometry has a smaller splitting between the two peaks and 
does not agree as well with the data.  The theory thus favors the
T$_d$ geometry for the ground state.  This is not the predicted ground
state in ref. \cite{bo93}, but since the calculated energy difference 
between geometries is only 0.08 eV, the theoretical ordering is
uncertain.  For the Ag$_9^+$ cluster, we used the geometry (I) of
ref. \cite{bo93}, the predicted ground state of the cluster in their
most detailed calculation.  The comparison between theory and experiment
\cite{ti93} is shown in Fig.~6.  The peak at 4 eV is reproduced in
position; its theoretical width is somewhat broadened due to the
lower geometric symmetry of the 9-atom cluster.

We next turn to the integrated absorption strength.  The strength
function $f_E$ is
shown in Fig.~5 for Ag$_8$ in the T$_d$ and D$_{2d}$ geometries;
the results for  Ag$_9^+$ are shown in Fig.~7.
The sharp modes below 5 eV are predicted to have only 25\% of the 
$s$-electron sum
rule.  This is slightly higher than the Mie theory prediction, which
perhaps can be attributed to the imperfect screening in a small
cluster. The same physics is responsible for the blue shift of the
excitation in small clusters.    Although the sharp
states are strongly screened, the integrated strength below 6 eV
is 3.9 electrons, about 50\% of the $s$-electron sum.  The integrated 
strength data is
compared with theory in Fig.~8, showing all the trend with increasing
cluster size.  The integrated strength per $s$-electron has moderate
decrease with increasing cluster size; no trend is discernible
in the experimental data.  Beyond N=1, the experimentally measured strength
is substantially larger than theory predicts. The data of ref.~\cite{ti93} is 
about a factor
of two larger than theory, as may also be seen in Fig.~7.  However, it 
is difficult to assess the
errors in that measurement, and the data of ref.~\cite{fe93} is
not seriously out of disagreement in view of their assigned error
bars. From a theoretically point of view, it is difficult to avoid
the $d$-electron screening and the resulting strong reduction of the
strength.  We present in the next section a semianalytic argument on this point.

\section{Interpretation}

In this section we will analyze the $d$-electron contribution to the
TDLDA response from an atomic point of view.  In the TDLDA, the bound
electrons can be treated separately because they only interact
through the common mean field.  In particular, there are no Pauli
exclusions corrections when combining $s\rightarrow p$ and $d\rightarrow p$
transition strength.  To describe the response from an atomic point
of view, it is convenient to express it in terms of the 
dynamic polarizability $\alpha(\omega)$.  
We remind the reader that
it is related to the strength function $S(E)=df_E/dE$
by
\be
\alpha(\omega)={e^2\hbar^2\over m}\int_0^\infty dE {S(E)\over -\omega^2 + E^2}.
\ee
The data in Table I may be used to estimate the $d\rightarrow p$ 
polarizability function, but this would not include higher energy
contributions and the continuum
$d\rightarrow f$ transitions. Instead, we recomputed the atomic
silver response freezing the $s$-electron.
That procedure yielded
a polarizability function with values $\alpha( 0\, {\rm eV})=1.8$~\AA$^3$
and $\alpha( 4\, {\rm eV})= 2.1$~\AA$^3$.  We then fit this
to a convenient single-state resonance form,
\be
\alpha_d = {e^2 \hbar^2\over m} { f_d\over -\omega^2 + E_d^2}.
\ee
with fit parameters $f_d=1.89$ and $E_d=10.7$ eV, from which we can
analytically calculate the effects on the $s$-electron response.
Except for minor interaction terms the TDLDA response is equivalent to
the RPA, which we apply using the response formalism as in App. A of ref.
\cite{ya95}.  Note that the dipole response function
$\Pi$ is related to the polarizability $\alpha$ by $\Pi=\alpha/e^2$.
Alternatively, the same physics can be
derived using the dielectric functions, as was done in 
ref.~\cite{li93,kr95,se97}.  The formulations are equivalent 
provided the dielectric function and the polarizability satisfy the
Clausius-Mossotti relation. In the dipole response formalism,
it is convenient to represent the uncoupled response function as
a $2\times 2$ matrix, separating the free-electron and the polarizability
contributions.  The RPA response function is written as
\be
\Pi^{RPA} = (1,1)(1+{\bf \Pi}^0 {\bf V})^{-1}{\bf \Pi}^0(1,1)^{t}
\ee
where ${\bf \Pi}^0$ and ${\bf V}$ are the following $2\times 2$ matrices:
\be
{\bf \Pi}^0 = \left(\matrix{\Pi^0_{free} & 0 \cr
0 & N \alpha_d/e^2 } \right)
\ee
\be
{\bf V} = {e^2\over R^3} \left(\matrix{1 & 1 \cr
1 & 0 } \right)
\ee
Here $N$ is the number of atoms in the cluster, and $R$ is
the radius of the cluster.
The form for ${\bf \Pi}^0$ is obvious, with the free electron
response given by $\Pi^0_{free}= -\hbar^2 N /m\omega^2$. The
${\bf V}$ is more subtle.
The Coulomb interaction, represented by the long-range
dipole-dipole coupling $e^2 \vec r_1\cdot\vec r_2/R^3$ \cite{be94}, 
acts among the free electrons and between the free electrons and
the polarization charge, but not within the polarization charges---
separated dipoles have zero interaction after averaging over
angular orientations.  The algebra in eq.~(4) is easily carried out
to give
\be
\Pi^{RPA}= {N\hbar^2/m\left(1-\alpha_d/r_s^3(1+\omega^2/\omega_M^2)
\right)\over-\omega^2 + \omega_M^2 (1-\alpha_d/r_s^3)}
\ee
where $r_s=(V/N)^{1/3}$ and  $\omega_M$ is the free-electron resonance
frequency defined in the introduction.  The pole position of the
response gives the frequency with the polarization,
\be
\omega_M'= \sqrt{1-\alpha_d/r_s^3} \,\,\,\omega_M
\ee
 Taking $r_s=3.09$ and $\alpha_d$ from the atomic calculation,
we find the resonance shifted from 5.18 to 3.6 eV, i.e. exactly 
the value for the empirical Mie theory.  .
The strength is calculated from the energy times the residue of the pole
which yields 
\be
f = N \left(1-{\alpha_d\over r_s^3}\right)^2
\ee
Numerically, eq. (9) gives a factor of 4 reduction in 
the strength, consistent with the full TDLDA calculation for Ag$_8$
with the $s+d$ valence space.  We thus conclude that the $d$-polarization
effects can be quite simply understood in atomic terms.

\section{Acknowledgment}

We acknowledge very helpful discussions with P.G. Reinhard, particularly
in formulating Sect. 4. 
This work is supported in part by 
the Department of Energy  under Grant DE-FG-06-90ER40561, and by the
Grant-in-Aid for Scientific Research from the Ministry of Education, 
Science and Culture (Japan), No. 09740236. Numerical calculations were
performed on the FACOM VPP-500 supercomputer in the institute for
Solid State Physics, University of Tokyo, and on the NEC sx4 supercomputer
in the research center for nuclear physics (RCNP), Osaka University.

\begin{table}
\caption{Atomic properties of Ag in the TDLDA.  The ionization potential 
on the first row is calculated by the static energy difference of
the silver atom and the singly-charged ion. See the text for explanation
of the other entries.
}
\begin{tabular} {ccccc}
&spherical basis  &  lattice  &  experimental\\
\tableline
I.P. (eV)&   8.0     & 8.0 eV  & 7.75 eV\\
\tableline
$e_p-e_s$ (eV) &   3.9 eV & 3.9 eV &\\
$E_{p\bar s}$ (eV)& 4.07 eV& 4.13 eV& 3.74 eV\\
$f_{p\bar s}$ & 0.53 & 0.66 & 0.7\\
\tableline
$e_p-e_d$ (eV)& 7.2 eV& 7.1 eV &&\\
$E_{p\bar d}$ (eV)&7.6 eV& 7.7 eV &8.2 eV\\
\end{tabular}
\end{table}

\begin{figure}
  \begin{center}
    \leavevmode
    \parbox{0.9\textwidth}
           {\psfig{file=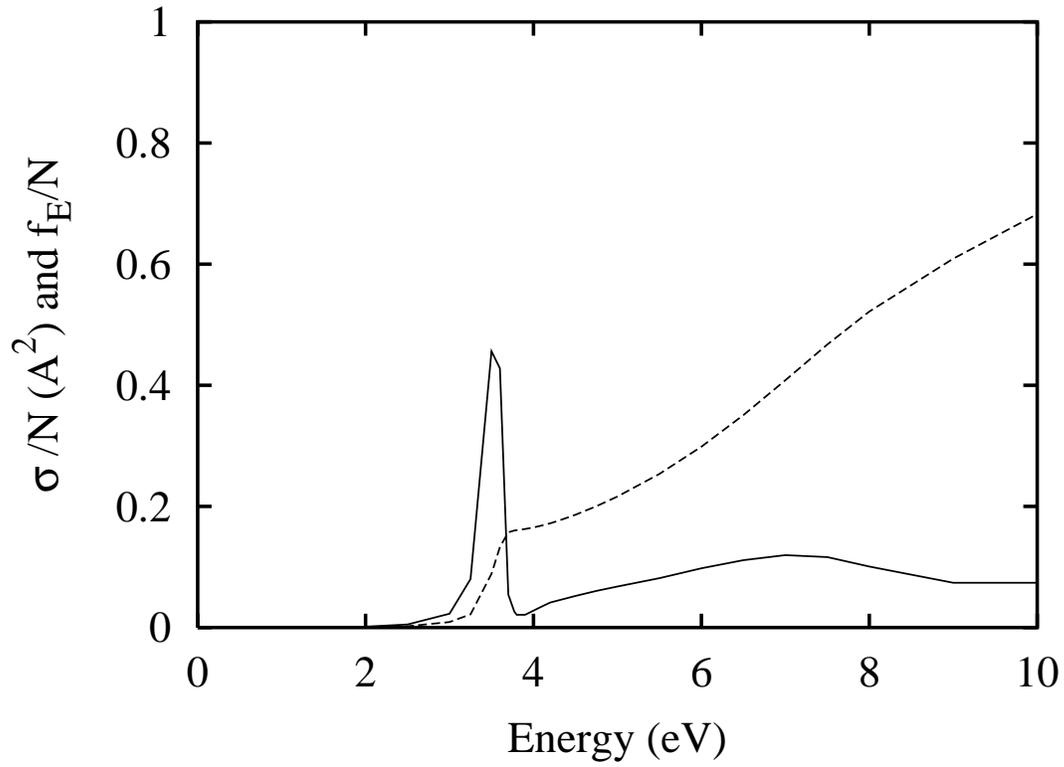,width=0.9\textwidth}}
  \end{center}
\caption{Mie theory of the optical absorption spectrum of
silver clusters.  Solid line is the absorption cross section
per atom, and the dashed line is the integrated oscillator
strength per atom.
}
\label{Mie}
\end{figure}

\begin{figure}
  \begin{center}
    \leavevmode
    \parbox{0.9\textwidth}
           {\psfig{file=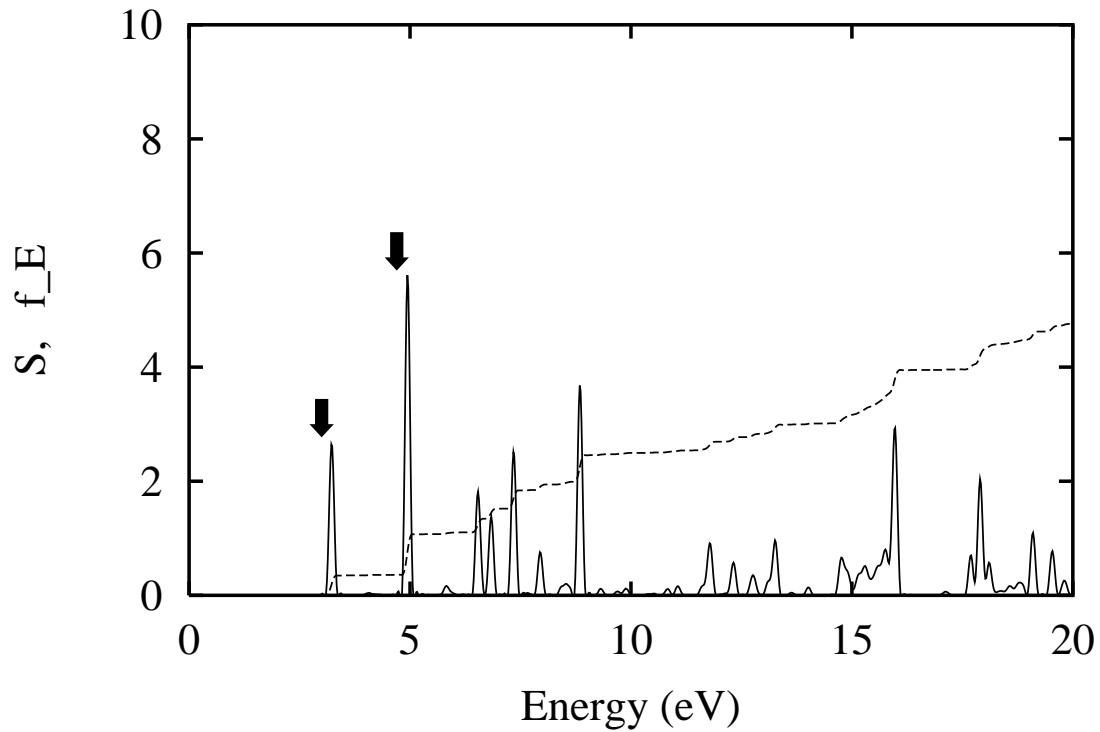,width=0.9\textwidth}}
  \end{center}
\caption{
Optical absorption spectrum of Ag$_2$.  Solid lines show the Fourier
transform of the TDLDA response; dashed lines showed the integrated
strength.  Arrows indicate peaks measured on clusters in
an argon matrix\protect\cite{fe93a}.
}
\label{ag2}
\end{figure}

\begin{figure}
  \begin{center}
    \leavevmode
    \parbox{0.9\textwidth}
           {\psfig{file=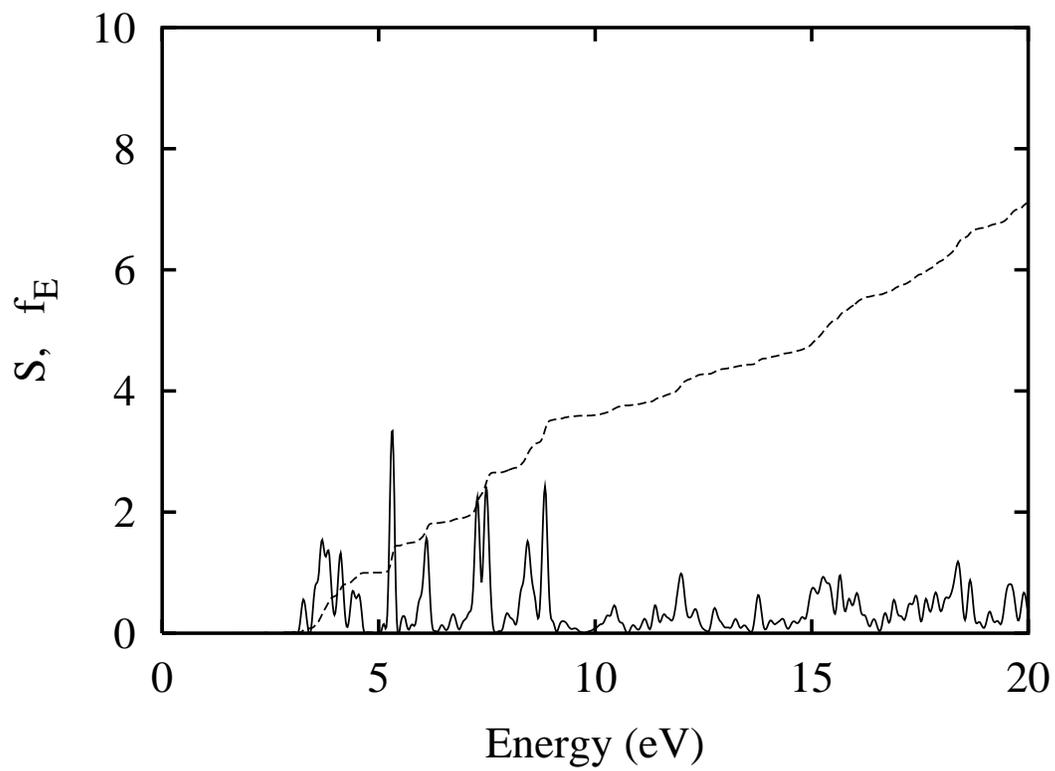,width=0.9\textwidth}}
  \end{center}
\caption{Optical absorption of Ag$_3$ in TDLDA.  
}
\end{figure}

\begin{figure}
  \begin{center}
    \leavevmode
    \parbox{0.9\textwidth}
           {\psfig{file=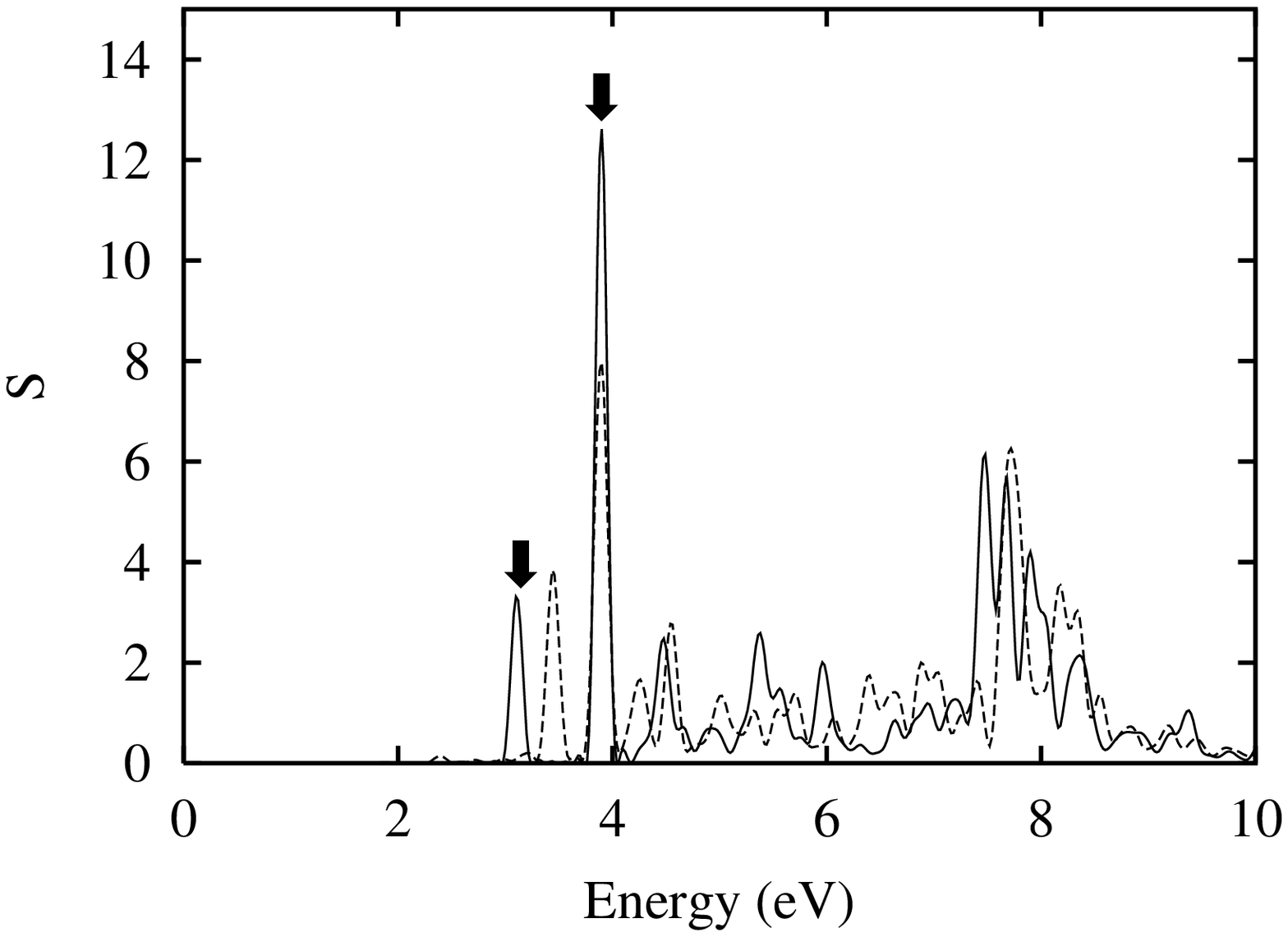,width=0.9\textwidth}}
  \end{center}
\caption{Optical Absorption spectrum of Ag$_8$.  Results for
two geometries are shown:  T$_d$ (solid) and D$_{2d}$ (dashed).
Arrows shows the position
of excitations observed in argon-matrix clusters, \protect\cite{fe93}.
}
\label{ag8s}
\end{figure}

\begin{figure}
  \begin{center}
    \leavevmode
    \parbox{0.9\textwidth}
           {\psfig{file=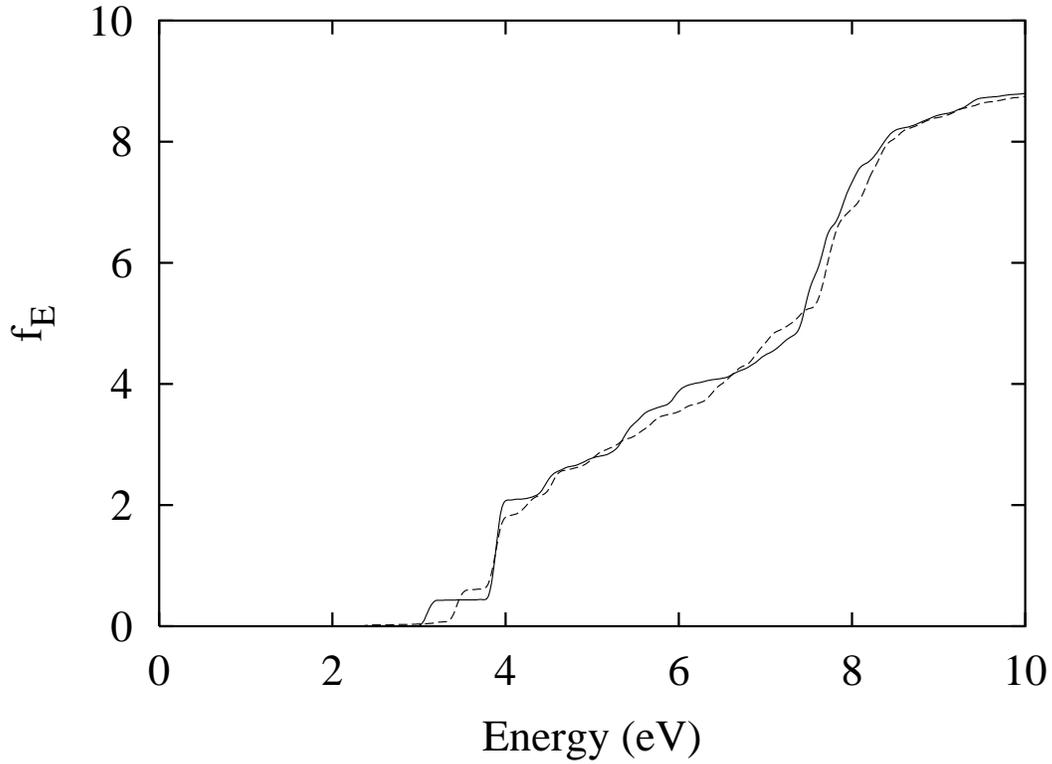,width=0.9\textwidth}}
  \end{center}
\caption{Integrated strength $f_E$ for Ag$_8$ in  
T$_d$ geometry (solid) and D$_2$ geometry (long dashed), and 
for Ag$_9^+$ (short dashed). 
}
\label{ag8f}
\end{figure}

\begin{figure}
  \begin{center}
    \leavevmode
    \parbox{0.9\textwidth}
           {\psfig{file=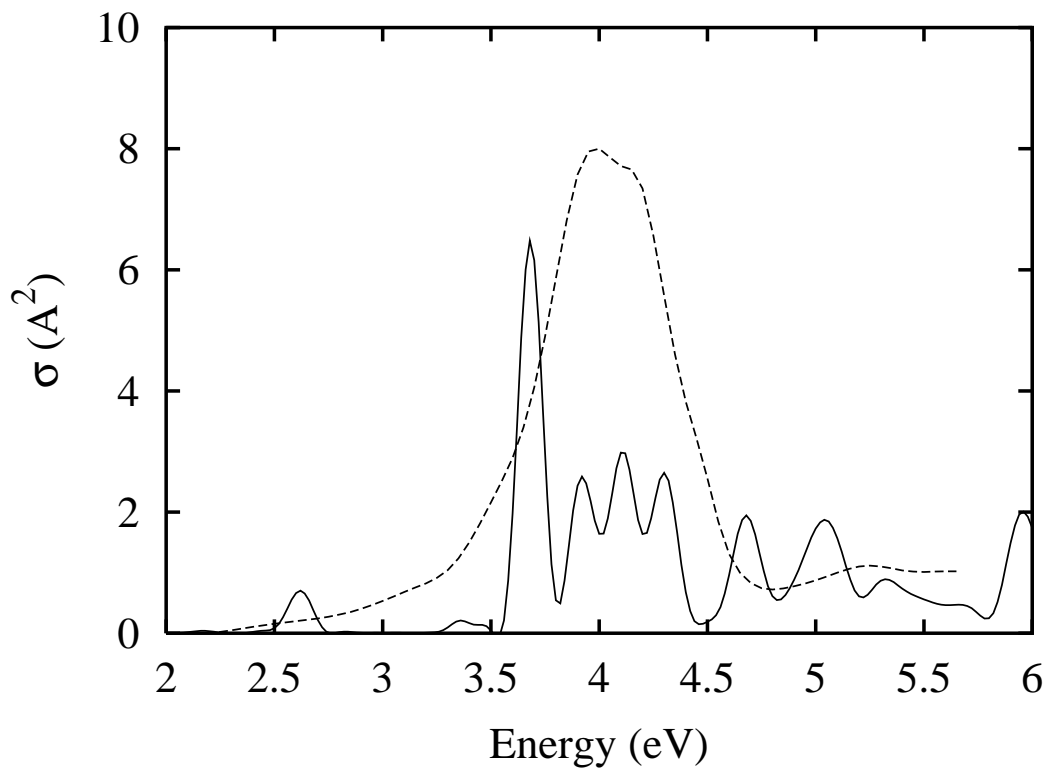,width=0.9\textwidth}}
  \end{center}
\caption{Absorption cross section in Ag$_9^+$ clusters:  TDLDA (solid);
experimental \protect\cite{ti93}(dashed).}
\label{ag9s}
\end{figure}

\begin{figure}
  \begin{center}
    \leavevmode
    \parbox{0.9\textwidth}
           {\psfig{file=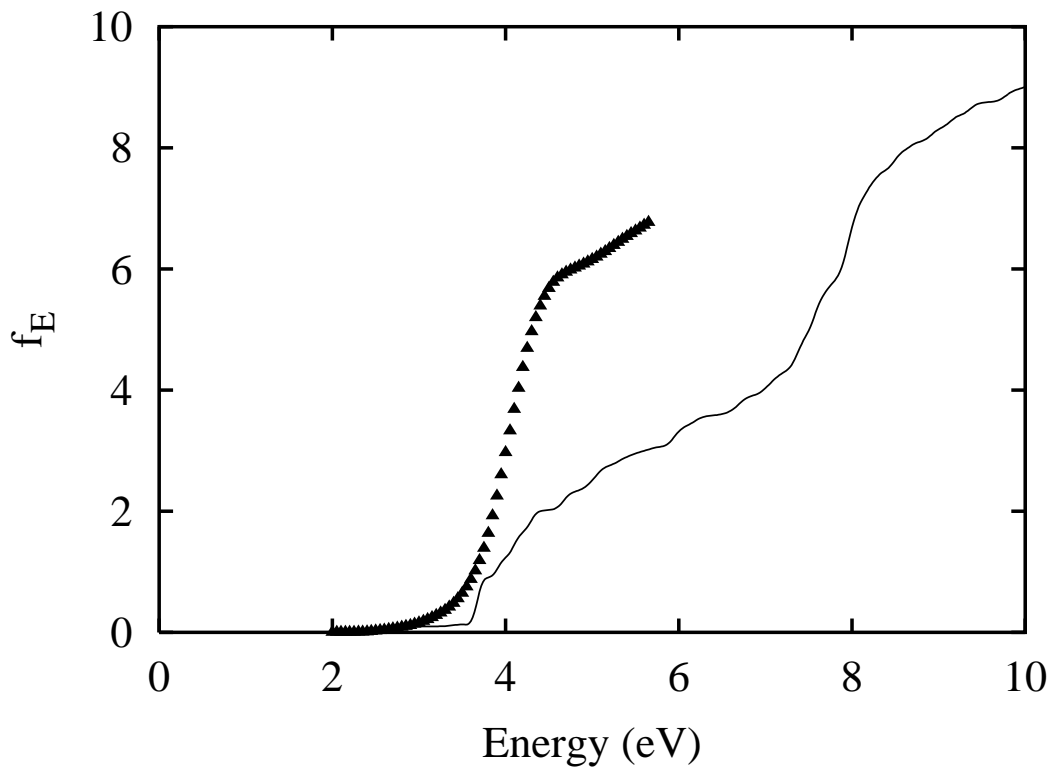,width=0.9\textwidth}}
  \end{center}
\caption{Integrated strength $f_E$ for Ag$_9^+$:  TDLDA (solid);
experimental \protect\cite{ti93} (triangles).}
\label{ag8f}
\end{figure}

\begin{figure}
  \begin{center}
    \leavevmode
    \parbox{0.9\textwidth}
           {\psfig{file=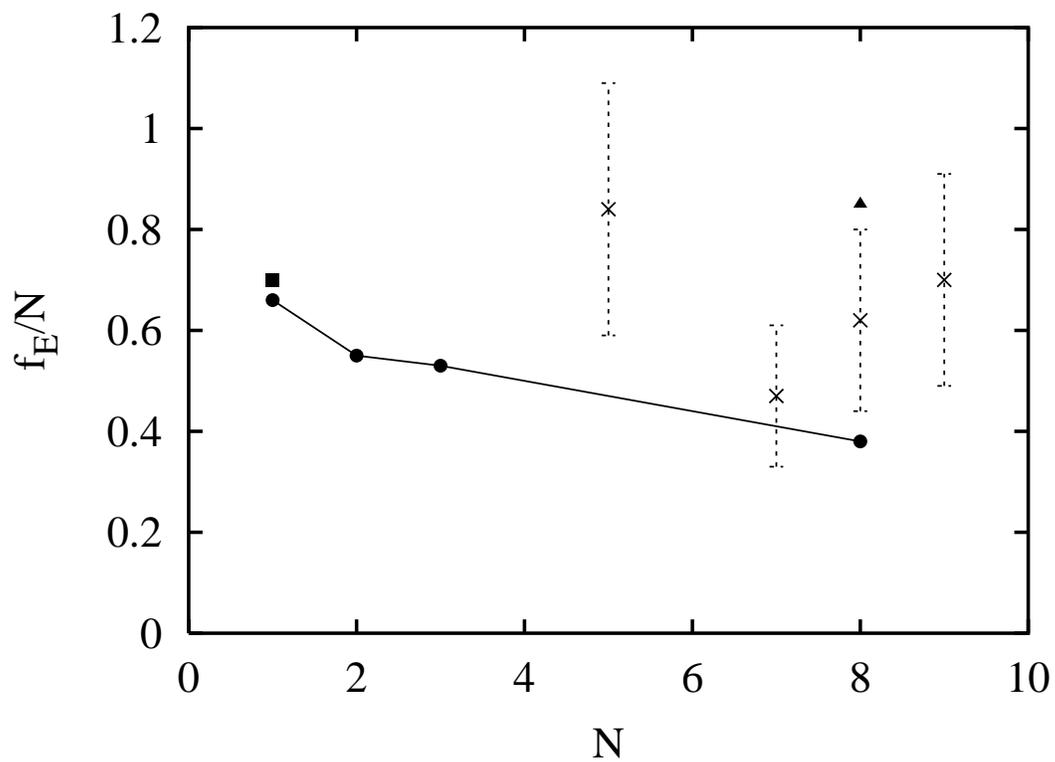,width=0.9\textwidth}}
  \end{center}
\caption{Integrated absorption strength below 6 eV as a function of
the number of $s$-electrons in the cluster.  TDLDA is given
by the circles connected with lines.  The source of the 
experimental data is: ref.~\protect\cite{pe63} (filled square);
ref.~\protect\cite{fe93} (crosses); ref~\protect\cite{ti93}
(triangle).
}
\label{f-sum}
\end{figure}

\end{document}